\documentclass[a4paper,11pt]{article}
\usepackage{pos}
\usepackage{Braket}
\usepackage{comment}

\begin{document}

\begin{flushright}
KEK-TH-2374
\end{flushright}

\title{A new phase in the Lorentzian type IIB matrix model and the emergence of continuous space-time}
\ShortTitle{A new phase in the Lorentzian type IIB matrix model}
\author*[a]{Mitsuaki Hirasawa}
\author[b]{Konstantinos N. Anagnostopoulos}
\author[c]{Takehiro Azuma}
\author[a]{Kohta Hatakeyama}
\author[d]{Yuta Ito}
\author[a, e]{Jun Nishimura}
\author[b]{Stratos Kovalkov Papadoudis}
\author[f]{Asato Tsuchiya}

\affiliation[a]{KEK Theory Center, High Energy Accelerator Research Organization,\\ 1-1 Oho, Tsukuba, Ibaraki 305-0801, Japan}
\affiliation[b]{Physics Department, School of Applied Mathematical and Physical Sciences, National Technical University,\\ Zografou Campus, GR-15780 Athens, Greece}
\affiliation[c]{Institute for Fundamental Sciences, Setsunan University,\\ 17-8 Ikeda Nakamachi, Neyagawa, Osaka, 572-8508, Japan}
\affiliation[d]{National Institute of Technology, Tokuyama College,\\ Gakuendai, Shunan, Yamaguchi 745-8585, Japan}
\affiliation[e]{Graduate University for Advanced Studies (SOKENDAI),\\ 1-1 Oho, Tsukuba, Ibaraki 305-0801, Japan}
\affiliation[f]{Department of Physics, Shizuoka University,\\ 836 Ohya, Suruga-ku, Shizuoka 422-8529, Japan}

\emailAdd{mitsuaki@post.kek.jp}
\emailAdd{konstant@mail.ntua.gr}
\emailAdd{azuma@mpg.setsunan.ac.jp}
\emailAdd{khat@post.kek.jp}
\emailAdd{y-itou@tokuyama.ac.jp}
\emailAdd{jnishi@post.kek.jp}
\emailAdd{sp10018@central.ntua.gr}
\emailAdd{tsuchiya.asato@shizuoka.ac.jp}

\abstract{The Lorentzian type IIB matrix model is a promising candidate for a non-perturbative formulation of superstring theory. In previous studies, Monte Carlo calculations provided interesting results indicating the spontaneous breaking of SO(9) to SO(3) and the emergence of (3+1)-dimensional space-time. However, an approximation was used to avoid the sign problem, which seemed to make the space-time structure singular. In this talk, we report our results obtained by using the complex Langevin method to overcome the sign problem instead of using this approximation. In particular, we discuss the emergence of continuous space-time in a new phase, which we discovered recently.}
\FullConference{The 38th International Symposium on Lattice Field Theory, LATTICE2021 26th-30th July, 2021
Zoom/Gather@Massachusetts Institute of Technology}
\maketitle
\section{Introduction}
Superstring theory is the most promising candidate for a unified theory of all interactions, including gravity.
The theory is consistently defined in ten-dimensional space-time, and the extra six dimensions need to be small enough to be consistent with current observations.
One mechanism that leads to phenomenologically acceptable scenarios is the compactification of the extra dimensions into small compact internal spaces. 
These scenarios have been investigated perturbatively on D-brane backgrounds and result in a vast number of vacua, leading to the so-called string landscape.
It is therefore interesting to see what happens when one includes nonperturbative effects and whether these play an essential role in determining the true
vacuum of the theory.

In 1996, the type IIB, or IKKT, matrix model was proposed as a nonperturbative formulation of superstring theory \cite{Ishibashi:1996xs}.
The model is obtained by dimensionally reducing the action of the ten-dimensional ${\cal N}=1$ Super Yang-Mills (SYM) to zero dimensions. The resulting matrix model
has maximal  ${\cal N}=2$ supersymmetry (SUSY) and the translations of the SUSY algebra is realized by the shifts $A_\mu\to A_\mu + \alpha_\mu \mathbf{1}$, $\mu=0,\ldots,9$.
The eigenvalues of the bosonic matrices $A_\mu$ are considered to define space-time; therefore, although space-time does not exist a priori in the model, it emerges
from the dynamics of the theory. The model has the potential to provide a nonperturbative mechanism for {\it dynamical} compactification of the extra dimensions in superstring theory.
Such a scenario has been shown to be realized in the model's Euclidean version. In this case, dynamical compactification is realized by the Spontaneous Symmetry Breaking (SSB) of
the SO(10) rotational symmetry down to SO(3), resulting in a three-dimensional macroscopic universe. 
This was shown by applying the Gaussian Expansion Method (GEM) \cite{Nishimura:2001sx,Kawai:2002jk,Aoyama:2006rk,Nishimura:2011xy}, 
which is a systematic expansion yielding nonperturbative information, and by Monte Carlo calculations \cite{Anagnostopoulos:2013xga,Anagnostopoulos:2015gua,Anagnostopoulos:2017gos,Anagnostopoulos:2020xai}.  

Those results provide a strong motivation to study the model in its original, Lorentzian version. A straightforward Monte Carlo calculation is not possible because the model has a strong complex
action problem. This led the authors in \cite{Kim:2011cr} to use an approximation that completely removes the complex action problem. They found that the model features a 
{\it dynamically} emerging continuous time, and that space is expanding from Planck scale to a macroscopic {\it three dimensional} universe. This happens by SSB the SO(9) rotational symmetry of
space down to SO(3). Following works showed that this expansion has
the potential to be phenomenologically viable, being exponential at short times and power like at late times \cite{Ito:2013qga,Ito:2013ywa,Ito:2015mxa,Ito:2015mem}. 
At late times, the dominant configurations can be approximated by classical solutions representing an expanding space that contain phenomenologically consistent matter content at low energies 
\cite{Kim:2011ts,Kim:2012mw,Chaney:2015ktw,Stern:2018wud,Steinacker:2010rh,Chatzistavrakidis:2010xi,Chatzistavrakidis:2011gs,Steinacker:2017bhb,Aoki:2010gv,Aoki:2014cya,Honda:2019bdi,
Hatakeyama:2019jyw,Steinacker:2021yxt}.

Recent work shows that the expansion is driven by singular configurations associated with the Pauli matrices, in which only two eigenvalues are large \cite{Aoki:2019tby}.
This behavior is due to the approximation used to avoid the complex action problem; therefore it becomes necessary to perform calculations in the full Lorentzian model.
In that case, the partition function is not well defined, and in \cite{Nishimura:2019qal} it was proposed to perform two independent world-sheet and target space Wick rotations, 
parameterized by two parameters $s$ and $k$, respectively, and then take an appropriate $s,k\to 0$ limit. Even after this deformation, the model suffers from a severe complex action problem,
which the authors overcame by using the Complex Langevin Method (CLM) \cite{Parisi:1983mgm,Klauder:1983sp}. 
The CLM has been known to yield wrong results in many interesting cases, but  by applying new techniques, and correct convergence criteria
\cite{Aarts:2009dg,Aarts:2009uq,Aarts:2011ax,Nishimura:2015pba,Nagata:2015uga,Nagata:2016vkn,Ito:2016efb}, 
it is possible to use the method successfully. Recently, the CLM  has been applied to the Euclidean IIB matrix model, yielding results that are in agreement 
with GEM calculations \cite{Anagnostopoulos:2017gos,Anagnostopoulos:2020xai}. 

This work applies the CLM to a simplified version of the IIB matrix model, where the fermionic degrees of freedom are quenched (bosonic IIB). The model is deformed by the Wick rotations 
mentioned above, and we are able to study $(s,k)$ deformations for $k=0$, $-1\le s \le 0$. The $s=-1$ model is the one studied in \cite{Kim:2011cr} and the
$s=0$ model is the bosonic Lorentzian IIB matrix model. The simulations are successful, and we observe a transition from the singular ``Pauli Matrix'' dominated phase to 
a new, continuous phase. We observe that the dominant configurations are continuous, non-expanding space-times. As in the case of the Euclidean model, we expect that 
SUSY plays an essential role in the SSB of the rotational symmetry of space. We expect to report results in this direction in future work.

\section{The Model}

The Lorentzian type IIB matrix model is given by
\begin{equation}
   Z=\int \mathcal{D}A \mathcal{D}\Psi e^{-S(A,\Psi)},
\end{equation}
\begin{equation}
   S(A,\Psi)=iN\beta \left\{ \frac{1}{4}{\rm Tr}\left[A_\mu, A_\nu\right]^2 + \frac{1}{2}{\rm Tr}\left(\Psi_\alpha (\mathcal{C} \Gamma^\mu)_{\alpha\beta} \left[A_\mu, \Psi_\beta \right] \right) \right\},
\label{action_Lor}
\end{equation}
where  $A_\mu$ and $\Psi_\alpha$ are $N\times N$ Hermitian matrices. 
$\Gamma^\mu$ and $\mathcal{C}$ are 10--dimensional gamma matrices and the charge conjugation matrix, respectively, which are obtained after the Weyl projection. 
The index $\mu$ runs from $0$ to $9$, and $\alpha$ runs from $1$ to $16$.
This model has an SO(9,1) Lorentz symmetry, under which $A_\mu$ and $\Psi_\alpha$ transform as a 10-dimensional Lorentz vector and a Majorana-Weyl spinor, respectively. 
Since the numerical cost for the evaluation of the fermionic part is very high, we neglect the fermionic contribution hereafter.
Namely, we omit the second term of eq.~(\ref{action_Lor}).

We consider the world-sheet and target space Wick rotations proposed in \cite{Nishimura:2019qal}, parameterized by the real parameters $(s,k)$
\begin{equation}
   \label{kk2}
   Z=\int \mathcal{D}A e^{-\bar{S}_{\rm b}(A)},
\end{equation}
\begin{equation}
   \label{kk1}
   \bar{S}_{\rm b}(A)=-iN\beta e^{is\pi/2}\left\{ \frac{1}{2}e^{-ik\pi}{\rm Tr}\left[A_0, A_i\right]^2 - \frac{1}{4}{\rm Tr}\left[A_i, A_j\right]^2 \right\}.
\end{equation}
The values of $(s,k)$ must be such that the real part of $\bar{S}_{\rm b}(A)$ is positive. 
When  $(s,k)=(0,0)$ we have the Lorentzian bosonic version of the IIB matrix model, whereas when $(s,k)=(1,1)$ we obtain the Euclidean bosonic version of the IIB matrix model\footnote{In 
\cite{Hatakeyama:2021}, we discuss the relation between the Euclidean and Lorentzian versions of type IIB matrix model.}.
For simplicity, we refer to those models as the ``Lorentzian'' and the ``Euclidean'' models.
The case  $(s,k) = (-1, 0)$ corresponds to the model first studied in \cite{Kim:2011cr}. 
Therefore, the model defined in (\ref{kk1}) interpolates continuously between these three cases.

We also introduce the IR constraints
\begin{equation}
\label{k2}
\frac{1}{N}{\rm Tr}\left(A_0\right)^2 = \kappa\, , \qquad \frac{1}{N}{\rm Tr}\left(A_i\right)^2 = 1\, ,
\end{equation}
in order to reduce the fluctuations of the zero modes in eq. (\ref{kk2}). 

One of the non-trivial features of the model is the emergence of continuous time from its dynamics.
If we choose the SU($N$) basis which diagonalizes the $A_0$ such that
\begin{equation}
\begin{split}
A_0 = {\rm diag}\left( \alpha_1, \alpha_2, ..., \alpha_N \right)\,,\qquad
\alpha_1 \le \alpha_2 \le \cdots \le \alpha_N,\label{k1}
\end{split}
\end{equation}
then, in the model studied in \cite{Kim:2011cr}, the spatial matrices exhibit a  band-diagonal structure, where $|(A_i)_{\nu+p,\nu+q}| \ll 1$ for $p^2+q^2 < n^2$.
We call $n$ the band width, and we define the $n\times n$ block matrices
\begin{equation}
\left(\bar{A_i}\right)_{ab}(t) = \left(A_i\right)_{\nu+a,\nu+b},
\end{equation}
where $t$ is defined by
\begin{equation}
t = \sum_{a=1}^n \left(A_0\right)_{\nu+a, \nu+a} = \sum_{a=1}^n \alpha_{\nu+a}.
\end{equation}
It is natural to consider that the matrices $\bar{A}_i(t)$ represent (fuzzy) space at time $t$.

\section{The Complex Langevin Method}
The complex Langevin method (CLM) was introduced by Klauder and Parisi independently as one of the methods to overcome the notorious sign problem \cite{Parisi:1983mgm, Klauder:1983sp}.
The idea of the CLM is complexifying the dynamical variable to evaluate the complex Boltzmann weight correctly.
In this section, we briefly explain the CLM.

Consider the case of a complex-valued action $S(x)\in\mathbb{C}$ with a set of real-valued dynamical variables $x\in\mathbb{R}^n$.
The partition function is given by
\begin{equation}
\label{k4}
Z=\int dx\ e^{-S(x)}.
\end{equation}
Since $e^{-S(x)}$ is complex, in most interesting cases, the usual Monte Carlo methods are not applicable.
In the CLM, we complexify the dynamical variables as
\begin{equation}
x\in\mathbb{R}^n \to z\in\mathbb{C}^n.
\end{equation}
Then the partition function is given by
\begin{equation}
Z=\int dz\ e^{-S(z)},
\end{equation}
and the complex Langevin equation is obtained as
\begin{equation}
\label{k3}
\frac{dz_i}{dt_{\rm L}} = -\frac{\partial S}{\partial z_i} + \eta_i(t_{\rm L}),
\end{equation}
where $t_{\rm L}$ is the (fictitious) Langevin time, the $\eta_i(t_{\rm L})$ can be chosen to be real Gaussian noise, which satisfies
\begin{equation}
\Braket{\eta_i(t_{\rm L})\eta_j(t_{\rm L}^\prime)}_\eta=2\delta_{i,j}\delta_{t_{\rm L},t_{\rm L}^\prime},
\end{equation}
and $\frac{\partial S(z)}{\partial z_i}$ is the drift term. 
The drift term $\frac{\partial S(z)}{\partial z_i}$ is obtained from  $\frac{\partial S(x)}{\partial x_i}$ by analytic continuation.

It is known that the CLM might yield wrong results.
Fortunately, a criterion for the solutions of  (\ref{k3}) to give the expectation values of the model defined by the partition function
(\ref{k4}) was proposed quite recently \cite{Aarts:2009uq, Nagata:2016vkn}.
The criterion is that the distribution of the drift term values should be suppressed exponentially or faster for a large magnitude of its values.
The assumption of holomorphicity is crucial for the correctness of the method.

The first step is to make a change of variables to realize the order of the eigenvalues of $A_0$ \cite{Nishimura:2019qal}.
We introduce new variables $\tau_i$ as
\begin{equation}
\alpha_a = \sum_{i=1}^{a-1} e^{\tau_i},
\end{equation}
and we treat the $\tau_i$ as the dynamical variables, which are complexified in the CLM. Then the order (\ref{k1}) is automatically realized.

Next we apply the CLM to the model.
The complex Langevin equation is the following:
\begin{equation}
\begin{split}
\frac{d\tau_a}{dt_{\rm L}} &= -\frac{\partial \bar{S}_{\rm eff}}{\partial \tau_a} + \eta_{a}(t_{\rm L}), \\
\frac{d(A_i)_{ab}}{dt_{\rm L}} &= -\frac{\partial \bar{S}_{\rm eff}}{\partial (A_i)_{ba}} + (\eta_i)_{ab}(t_{\rm L}),
\end{split}
\end{equation}
where $t_{\rm L}$ is the Langevin time, and $\eta$ is the Gaussian noise. The matrices $(\eta_i)_{ab}(t_{\rm L})$ are chosen to be Hermitian.
The effective action $\bar{S}_{\rm eff}$ is obtained from $\bar{S}_{\rm b}$, by adding the appropriate gauge fixing and change of variable terms, as well as terms that enforce the constraints (\ref{k2}).
In the original model, the drift terms $\frac{\partial \bar{S}_{\rm eff}}{\partial \tau_a}$ and $\frac{\partial \bar{S}_{\rm eff}}{\partial (A_i)_{ba}}$ are defined for real variables $\tau_{a}$ and Hermitian matrices $(A_i)_{ba}$. In the CLM, we perform an analytic continuation by taking complex ${\tau_a}$ and  general complex matrices $(A_i)_{ba}$.

\section{Results}
First we show the result for $(s, k) = (-1, 0)$ case, which corresponds to the approximation used in \cite{Kim:2012mw,Aoki:2019tby}.
Then we try to approach $(s, k) = (0, 0)$, which corresponds to the Lorentzian model.

\subsection{$(s, k) = (-1, 0)$ case}
\begin{figure}
\centering{}
\includegraphics[scale=0.45]{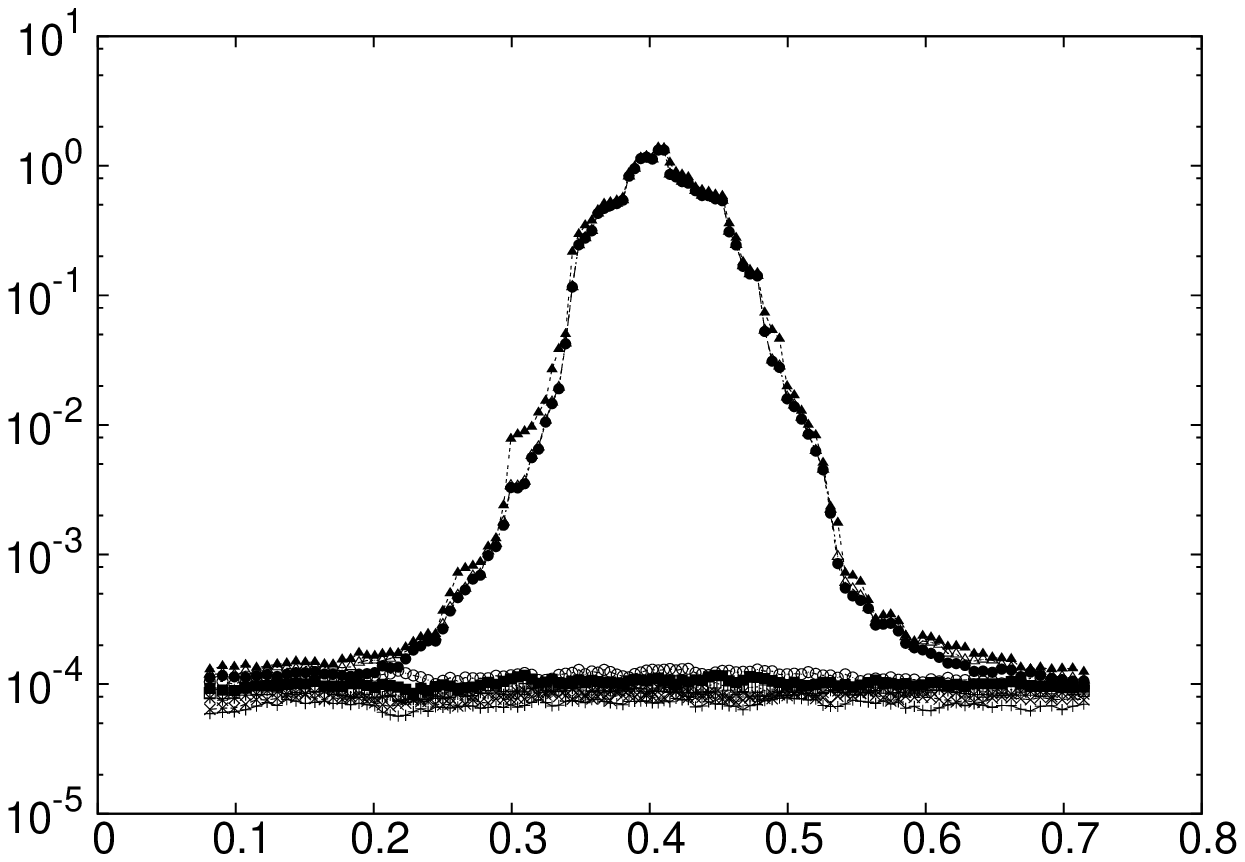}
\includegraphics[scale=0.45]{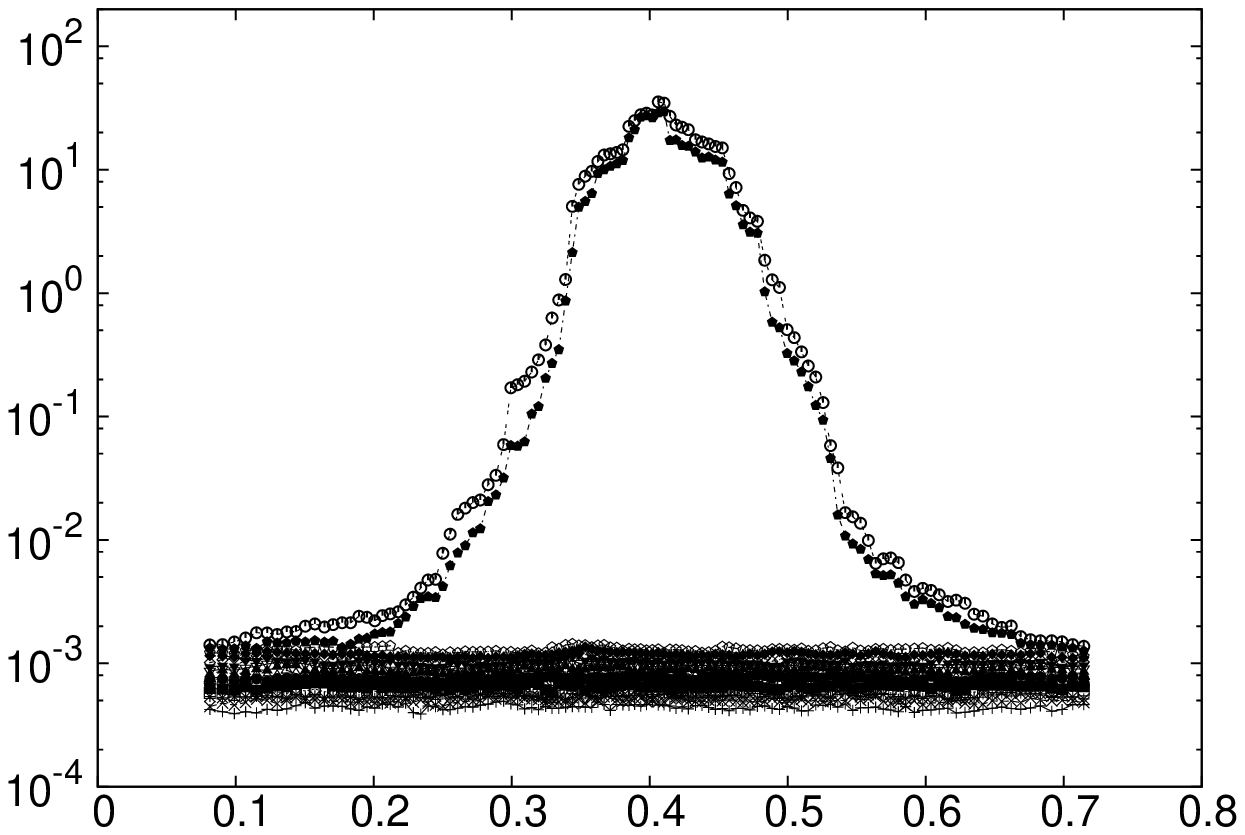}
\caption{Results for the bosonic model with $(s, k) = (-1, 0),\ N=128,\ \beta=8.0,\ \kappa = 0.04$.
(Left) The eigenvalues of $T_{ij}(t)$ against time $t$. We find that only 3 out of 9 eigenvalues grow as time increases.
(Right) The eigenvalues of $Q(t)$ against time $(t)$. Only 2 eigenvalues grow as time increases.}
\label{fig:s-1k0}
\end{figure}

In order to study the SSB of the the SO(9) spatial symmetry, we define ``the moment of inertia tensor" as
\begin{equation}
T_{ij}(t) \equiv {\rm tr}\left(\bar{X}_i(t)\bar{X}_j(t)  \right),
\end{equation}
where $\bar{X}_i(t)$ is defined as
\begin{equation}
\bar{X}_i(t) = \frac{1}{2}\left(\bar{A}_i(t) + \bar{A}^\dagger_i(t) \right),
\end{equation}
and ``${\rm tr}$" means the trace over the block matrices.
When the space is SO(9) symmetric, the 9 eigenvalues have the same large--$N$ limit.
In Figure~{\ref{fig:s-1k0}} (Left), we plot the eigenvalues of $T_{ij}(t)$ against time $t$. At early times, the 9 eigenvalues are almost the same. On the other hand, 3 out of 9 eigenvalues start to grow at some time. This behavior indicates that SO(9) symmetry is spontaneously broken at this time, and a macroscopic SO(3) symmetric space emerges.

Next, to study the structure of the space, we define $Q(t)$ as
\begin{equation}
Q(t) \equiv \sum_{i=1}^9 \left(\bar{X}_i(t)\right)^2,
\end{equation}
whose eigenvalues describe how space spreads into the radial direction. If the space is smooth, the spectrum of the eigenvalues is continuous.
In Figure~{\ref{fig:s-1k0}} (Right), we plot the eigenvalues of $Q(t)$ against time $t$.
Only 2 eigenvalues start to grow at some point, which corresponds to when the SO(3) symmetric space starts to expand.
This behavior means that the expansion of space is realized by only two isolated points at large distance.
Therefore, space is not continuous \cite{Aoki:2019tby}.

\subsection{approaching $(s, k) = (0, 0)$}
\begin{figure}
\centering{}
\includegraphics[scale=0.35]{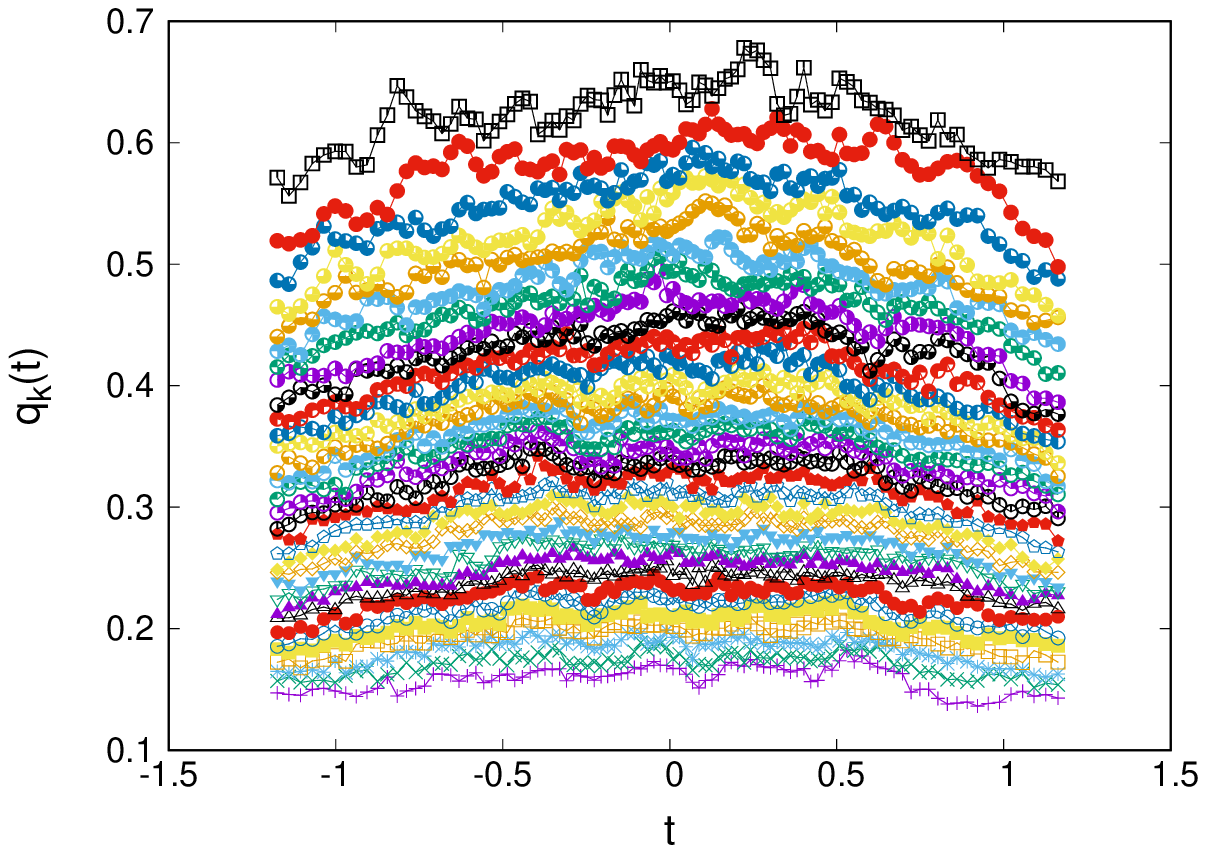}
\includegraphics[scale=0.35]{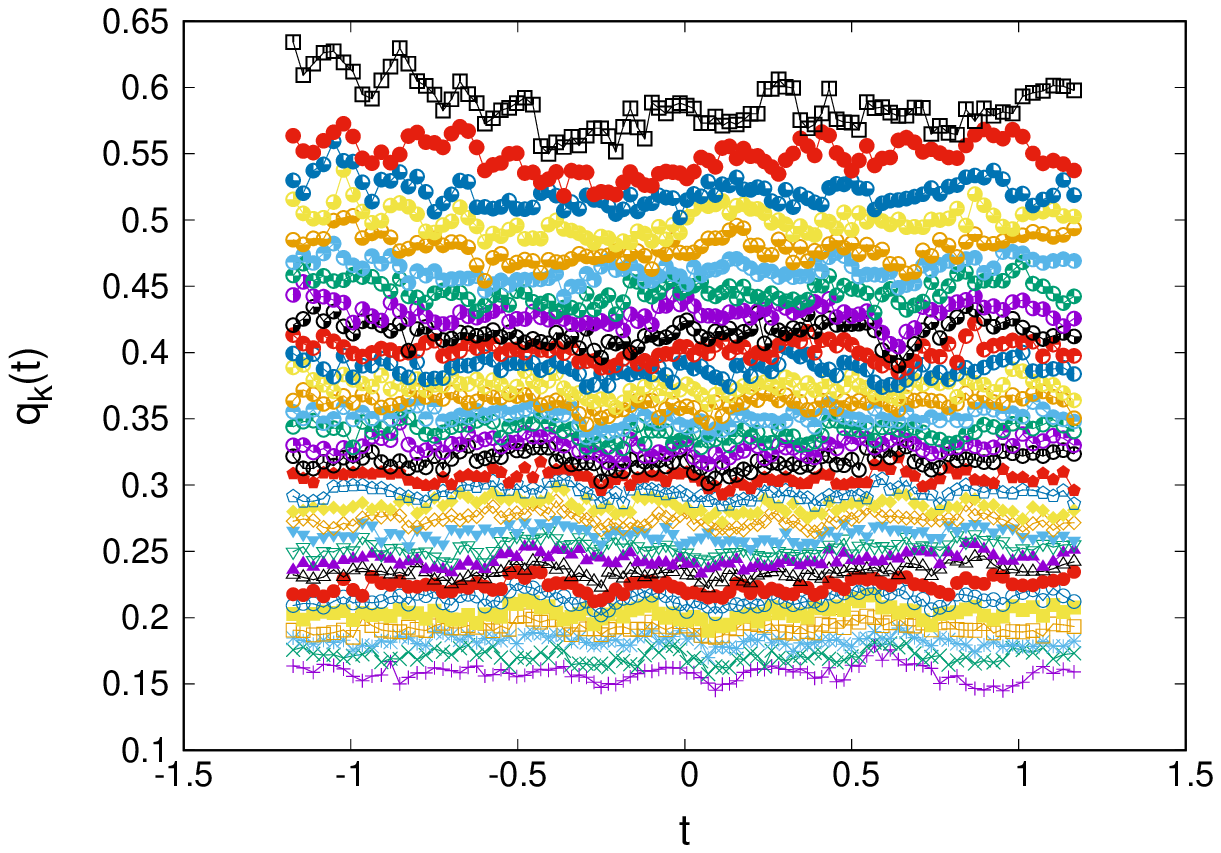}
\includegraphics[scale=0.35]{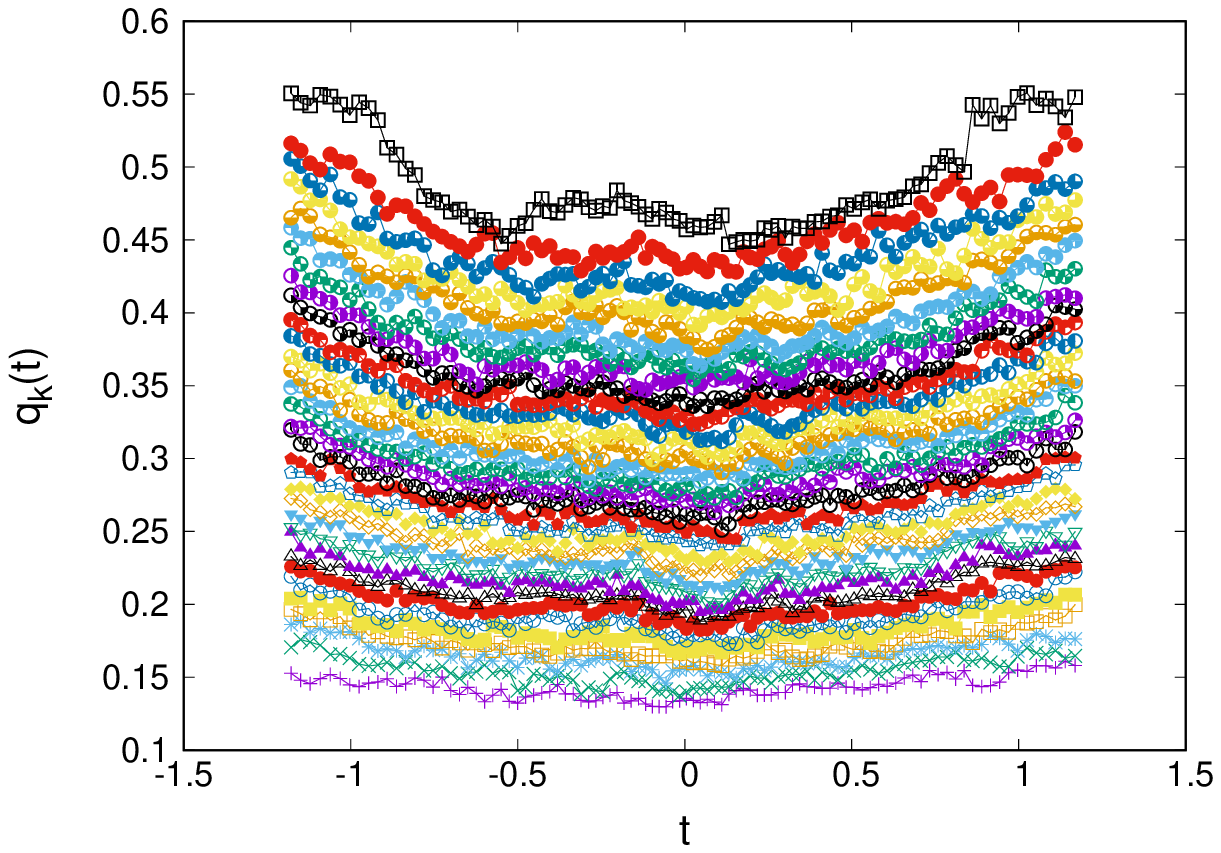}
\caption{The eigenvalues of $Q(t)$ are plotted against time $t$.
(Left) The $(s, k) = (-0.8, 0)$ case.
(Center) The $(s, k) = (-0.6, 0)$ case.
(Right) The $(s, k) = (0, 0)$ case.
The other parameters are common, namely $N=128,\ \beta=2.5,\ \kappa=0.8$.
}
\label{fig:app_lor}
\end{figure}

We try to approach $(s, k) = (0, 0)$ by using the CLM.
We fix $k=0$ and do the simulations for various values of $s$.
In Figure~\ref{fig:app_lor}, we plot results for 3 different values of $s$.
The (Left), (Center), and (Right) plots correspond to $s=-0.8,\ -0.6,\ {\rm and}\ 0$ respectively.
The expansion behavior changes as $s$ approaches $0$.
When $s\approx -0.8$, a bell-shaped expansion appears.
If $s$ gets closer to $0$, at some point the expansion disappears ($s\approx -0.6$), and finally a parabola-shaped  expansion appears at $s=0$.
This parabola-shaped expansion is consistent with typical classical solutions \cite{Hatakeyama:2019jyw}.

\begin{figure}
\centering{}
\includegraphics[scale=0.50]{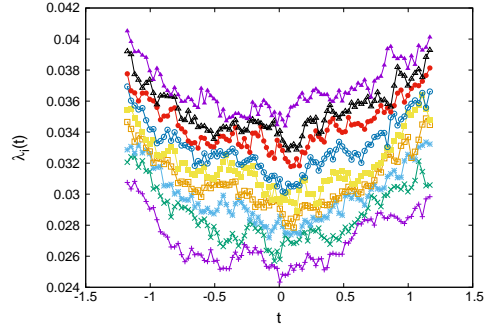}
\caption{
Result for $(s, k)=(0, 0),\ N=128,\ \beta=2.5,$ and $\kappa=0.8$.
The eigenvalues $\lambda_i(t)$ of $T_{ij}(t)$ are plotted against time $t$.
}
\label{fig:T_s0k0}
\end{figure}

In Figure~\ref{fig:T_s0k0}, we plot the eigenvalues of $T_{ij}(t)$ against time $t$ for the $(s, k) = (0, 0)$ case.
We observe that the 9 eigenvalues do not indicate an SSB pattern, even at late times.
Therefore the SSB of SO(9) does not occur in this case.

\section{Summary and discussion}
We applied the complex Langevin method (CLM) to the bosonic version of the type IIB matrix model to overcome the sign problem.
As in \cite{Nishimura:2019qal}, we introduced the two parameters $s$ and $k$, related to the Wick rotation on the world sheet and target space-time, respectively.
The CLM was applied successfully, and we could simulate the model even at $s=k=0$.
In this work, we explored the $k=0$, $-1\leq s \leq 0$ line, and we found a new phase in which the continuous space emerges as $s$ approaches $0$.
In this phase, the expansion behavior changes and is consistent with the one found for the classical solutions obtained in \cite{Hatakeyama:2019jyw}.
However, the SSB of the SO(9) rotational symmetry is not observed.

In \cite{Hatakeyama:2021}, we discuss the relation between the Lorentzian ($s=k=0$) and the Euclidean ($s=k=1$) case.
We have found that the behavior at the new phase is equivalent to the behavior obtained in the Euclidean model.
This paper also discusses possible scenarios for emergent Lorentzian space-time at late times.

It is known that supersymmetry plays a central role in realizing the SSB of the SO(10) to SO(3) rotational symmetry in the Euclidean model \cite{Anagnostopoulos:2020xai}, and by
neglecting the effects of the fermions, the SSB does not occur.
Therefore, we expect that supersymmetry will play an essential role in realizing the SSB of the SO(9) symmetry, leading to an expanding space and a promising matrix model cosmology.

\section*{Acknowledgment}
\noindent 
T. A., K. H., and A. T. were supported in part by Grant-in-Aid (Nos. 17K05425, 19J10002, and 18K03614, 21K03532, respectively) from Japan Society for the Promotion of Science. 
This research was supported by MEXT as “Program for Promoting Researches on the Supercomputer
Fugaku” (Simulation for basic science: from fundamental laws of particles to creation of nuclei) and
JICFuS.
The numerical computations were carried out on the K computer (Project ID : hp170229, hp180178), the Oakbridge-CX in University of Tokyo (Project ID : hp120281, hp200106), the PC clusters in the KEK Computing Research Center and the KEK Theory Center, and the XC40 at YITP in Kyoto University.
This work was also supported by computational time granted by the Greek Research and Technology Network (GRNET) in the National HPC facility ARIS, under the project IDs SUSYMM and SUSYMM2.

\bibliographystyle{JHEP}
\bibliography{bib.bib}
\end{document}